\documentclass[prl,aps,floatfix,showpacs,superscriptaddress,twocolumn]{revtex4}
\usepackage{graphicx,dcolumn,url}

\def \pbp              {\mbox{$\overline{p}p $}} 
\def \qqbar            {\mbox{$q\overline{q} $}} 
\def \shat             {\mbox{$\hat{s}$}} 
\def  \Et               {\mbox{$E_T$}}

\begin{document}

\title{
Search for $W$ and $Z$ Bosons in the Reaction $\pbp \rightarrow 2\ $ jets~$+~\gamma$ at 
$\sqrt{s}$ = $1.8$ TeV
}
\author{D.~Acosta}
\affiliation{University of Florida, Gainesville, Florida  32611}
\author{T.~Affolder}
\affiliation{University of California at Santa Barbara, Santa Barbara, 
California 93106}
\author{M.G.~Albrow}
\affiliation{Fermi National Accelerator Laboratory, Batavia, Illinois 60510}
\author{D.~Ambrose}
\affiliation{University of Pennsylvania, Philadelphia, Pennsylvania 19104}
\author{D.~Amidei}
\affiliation{University of Michigan, Ann Arbor, Michigan 48109}
\author{K.~Anikeev}
\affiliation{Massachusetts Institute of Technology, Cambridge, Massachusetts  
02139}
\author{J.~Antos}
\affiliation{Institute of Physics, Academia Sinica, Taipei, Taiwan 11529, 
Republic of China}
\author{G.~Apollinari}
\affiliation{Fermi National Accelerator Laboratory, Batavia, Illinois 60510}
\author{T.~Arisawa}
\affiliation{Waseda University, Tokyo 169, Japan}
\author{A.~Artikov}
\affiliation{Joint Institute for Nuclear Research, RU-141980 Dubna, Russia}
% ...
\collaboration{CDF Collaboration}
\noaffiliation

\date{\today}

\begin{abstract}

We present a study of the dijet invariant mass distribution for the reaction 
$\pbp \rightarrow 2\ $jets$+\gamma+X$, at a center of mass energy of $1.8$ TeV, 
using data collected by the CDF experiment.  We compare the data to predictions 
for the production of a photon with two jets, together 
with   the 
resonant processes $\pbp \rightarrow W/Z+\gamma+X$, in which the $W$ and $Z$ 
bosons decay hadronically.  A fit is made to the dijet invariant mass 
distribution combining the non-resonant background and resonant processes.  We 
use the result to establish a limit for the inclusive production cross section 
of $W/Z+\gamma$ with hadronic decay of the $W$ and $Z$ bosons.
\end{abstract}


\maketitle
 
\section{INTRODUCTION}
The production of final states containing $W$ or $Z$ bosons in association with a 
photon ($\gamma$) in $\pbp$ collisions at $\sqrt{s}=1.8$ TeV has been studied by 
the CDF and D$\O$ collaborations using event samples in which the $W$ and $Z$ 
bosons decay to leptons \cite{WZgamma}.  Identification of $W\gamma$ and 
$Z\gamma$ events in which the $W/Z$ decay hadronically is 
experimentally difficult because of large background from direct
production of $2$ jet + $\gamma$ and three-jet events. A study of the general kinematic 
characteristics of $2$ jet + $\gamma$ production has been  conducted 
using $16$ pb$^{-1}$ of data from 
CDF Run 1a, however the dijet invariant mass distribution 
was not investigated \cite{gam2jet}. 
\par
We present a search for evidence 
of $\gamma + W/Z \rightarrow \qqbar$ final states in $90$ pb$^{-1}$ of $\pbp$ data.
The data were collected with the CDF detector during Run 1b of the Tevatron. A search for
$2$ jet + $\gamma$ events was conducted within the data subset in which  
a photon candidate had electromagnetic transverse energy  greater than $23$ GeV.
The dijet invariant 
mass distribution was fit to a mixture of boson resonance decay and QCD background. 
\par
The methods established in this 
analysis \cite{marina} could prove useful for
identifying similar signals coming from $X \rightarrow jj$ decay which are embedded in
large QCD background.
\section{THE DETECTOR}
A comprehensive description of the Collider Detector at Fermilab (CDF) may be 
found in \cite{detector}.  We used a coordinate system with $z$ along the proton 
beam, azimuthal angle $\phi$, polar angle $\theta$, and pseudo-rapidity 
$\eta=-{\mathrm ln}\ {\mathrm tan}(\theta/2)$. The transverse energy of a particle (e.g. photon, electron,
jet)  was 
defined as $E_T = E \sin \theta$. The primary components relevant to this analysis 
were those that measure jet energies and positions, photon energies, positions and
profiles, and those 
that establish the $\pbp$ interaction vertex.
\par
The central tracking chamber (CTC) and vertex tracking chamber (VTX) were used to 
measure momenta and directions of the charged tracks 
associated with jets. The tracking chambers were located within $1.4$ T axial magnetic 
field. The CTC was a  drift chamber which provides space 
point information used to construct the trajectories of charged particles.  It covered 
a rapidity range of $|\eta| < 1.1$.  The VTX was a time projection chamber 
positioned between the beam pipe and the CTC that provided improved interaction 
vertex measurement with the extrapolation of tracks reconstructed in the CTC.
\par
Scintillator-based electromagnetic (CEM) and hadronic (CHA) calorimeters in the 
central region ($|\eta| \le 1.1$) were arranged in projective towers of size 
$\Delta\eta \times \Delta\phi \approx 0.1 \times 0.26$.  The end-wall hadronic 
calorimeter (WHA) and the end-plug electromagnetic (PEM) and  
hadronic (PHA) calorimeters extended the rapidity coverage out to $|\eta| \le 
2.4$.
\par
Two additional detector elements were used for photon identification.  The 
central strip chambers (CES) were multi-wire proportional chambers with segmented 
cathode strips. The CES was positioned near the shower maximum of the 
central electromagnetic detector.  The anode wires measured $\phi$ and the 
cathode strips measure $\eta$ for showers in the CEM.  The central preradiator 
(CPR) was a set of multi-wire proportional chambers, positioned between the 
magnet solenoid and the CEM. It was used to measure the electromagnetic shower pulse heights of 
electron-positron pairs from photons converting in a solenoid of 
thickness $1.1$ radiation lengths. The CES and the CPR detector systems
 provided discrimination between single 
photons and multi-photon showers arising from $\pi^0$ and $\eta$ decays.
\section{EVENT SELECTION}
Events with photon candidates were selected using a trigger which required a 
high threshold of energy deposited in the central electromagnetic calorimeter.
Photon clusters were reconstructed by combining the energy from neighboring CEM 
cells with a seed cell having energy above a threshold of $3$ GeV.  A photon candidate 
was required to have clustered energy $E_T >  23$ GeV.
\par
The trigger acceptance for photon candidates has been measured as a function of the photon
transverse momentum \cite{dana1}. The acceptance 
plateaus for the photons with $E_T\ge 30$ GeV at $0.970\pm0.006$.
\par
In the offline data analysis, the measured photon energies were corrected using an
 algorithm  taking into account variations of CEM response within
a cell, cell-to-cell energy sharing, and a global transverse energy scale \cite{cemfix}.  A photon 
candidate was required to be isolated, 
with less than $15\%$ additional energy ($E_{CEM} + E_{CHA}$) within a cone of 
radius $\Delta R = \sqrt{\Delta \phi^2 + \Delta \eta^2}<0.4$ centered on the photon direction.
In order to reject electrons, a tracking isolation requirement was imposed by summing the transverse 
momentum of tracks within the $\Delta R<0.4$ cone around the photon
direction.  The sum of $p_T$ of the tracks was required to be 
less than $2$ GeV/c.
\par
The ratio of the energy deposited in the hadronic calorimeter ($E_{HAD}$) to the energy 
deposited in the electromagnetic calorimeter ($E_{CEM}$) for a photon candidate passing the 
above cuts must satisfy the requirement
\[\frac{E_{HAD}}{E_{CEM}} \le 0.055 + 0.00045 E(\gamma),\]
with $E(\gamma)$ in GeV.
Typically, for a photon, the above ratio is less than $10\%$. A 
lateral sharing parameter for CEM clusters measures the spread of energy over 
calorimeter cells adjacent to the seed cell and selects photons based upon the spread 
expected from measurements in an electron test beam.
 \par
The photons had to pass fiducial cuts that require sufficient shower containment in the CES chambers.  The
 highest energy strip cluster and the highest energy wire cluster were chosen to estimate the position 
and transverse profile of a photon candidate.  Any additional clusters in the CES within the boundaries 
of the calorimeter energy cluster had to have energy less than $2.39$ GeV $+ 0.01$ $E_T(\gamma)$  GeV.
\par
Multi-photon backgrounds from $\pi^0$ and $\eta$ decays are suppressed by 
requiring the transverse profile of the shower energy observed in
      the CES to match to a reference profile measured using an electron test
beam. A partial separation of direct photons from multi-photon backgrounds was made based on the 
quality of the shower shape agreement.  Additional details of the photon cuts may be found in~\cite{gamanal}. 
\par
The energy from identified photons is removed from the calorimeter cells and the remaining energy is 
clustered to reconstruct jets.  Hadronic jets are identified using a cone clustering algorithm~\cite{jetclu}.
The energies of jets clustered with a cone size of $\Delta R<0.7$ were corrected for 
calorimeter nonlinearity and 
cell-to-cell non-uniformity. A correction takes into account energy deposition from particles
 not associated with the parent parton falling within the jet clustering cone, and an 
out-of-cone correction for the underlying event accounts for energy from the jet which is 
radiated outside the jet clustering cone.
\par
The sample of events with exactly two jets having $E_T$ greater than $15$~GeV and 
$|\eta|\le 1.1$ is used. The restricted rapidity range insures that the reconstruction efficiency
for tracks associated with these jets is uniform. Any additional jets present in the event 
within $|\eta|<2.4$ must have $E_T\leq 10$~GeV. Each of the two central jets must include at least 
one well reconstructed track, which is used to calculate the vertex position.  The vertices determined 
from the tracking information from the two jets in an event must lie within $10$ cm of each other. This 
cut removes instances of jets from different $p\bar{p}$ collisions during the same beam-crossing.  
The interaction vertex is taken to be the average of the two jet vertices and 
must lie within $60$~cm of the center of the detector.  This vertex, together with the information 
from the CES, is used to define the photon trajectory.
\par
A total of approximately $2.7$ million triggered events was recorded. A  sample of $9493$ events satisfied 
all photon and jet selection criteria.  The photon $\Et$ distribution is shown in Figure $1$.  Figure $2$ 
shows 
the $E_T$ and $\eta$ distributions for the two leading jets in the event, ordered by $E_T$.  The 
first jet has average $E_T$ of $39$ GeV, while the second jet has average $E_T$ of $23$ GeV.  A 
plot of the dijet invariant mass distribution is shown in Figure $3$. The 
dijet invariant mass $m_{jj}$ is defined as $\sqrt{(E_1 + E_2)^2 - (\vec{P_1} +\vec{P_2})^2}$, 
where $E_i$ and $\vec P_i$ are the energies and momenta of the two leading jets. 
\begin{figure}[htb!]
\begin{center}
\includegraphics[width=\columnwidth]{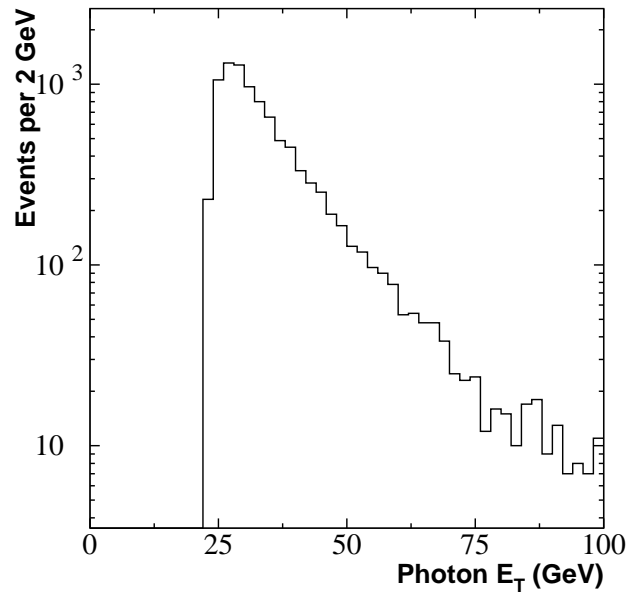}
\caption{Photon $E_T$ Distribution from $\gamma + 2$ jet events.
\label{Figure 1}}
\end{center}
\end{figure}
\begin{figure}[htb!]
\begin{center}
\includegraphics[width=\columnwidth]{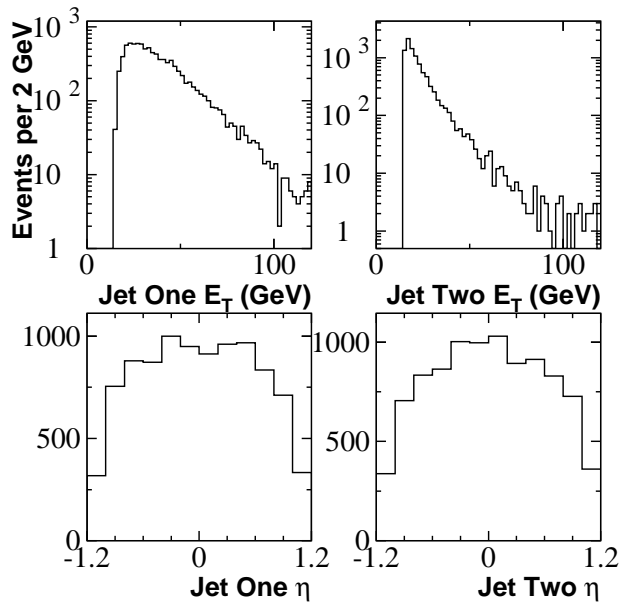}
\caption{The $E_T$ and $\eta$ distributions for the two leading jets.  
\label{Figure 2}}
\end{center}
\end{figure}
\begin{figure}[htb!]
\begin{center}
\includegraphics[width=\columnwidth]{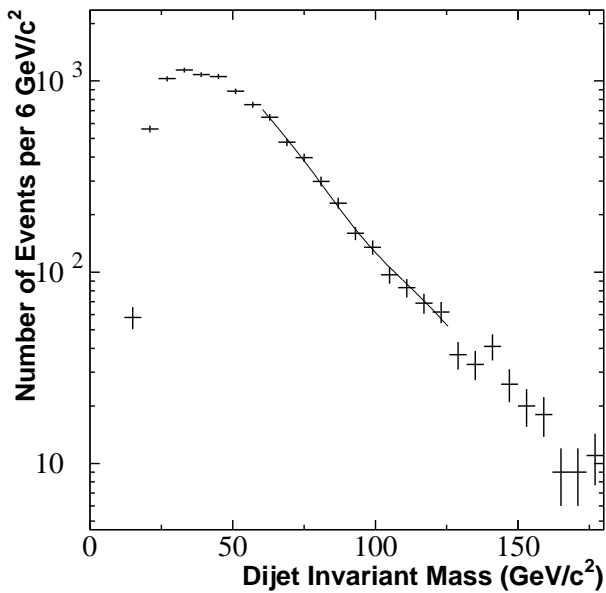}
\caption{ Dijet mass spectrum from $2$ jet $+ \gamma$ data. There are $2656$ 
events in the $m_{jj}$ region from $60$ to $126$ GeV/$c^2$. The solid fit corresponds to the 
background $+$ signal fit 
as described in Section $6$.
\label{Figure 3}}
\end{center}
\end{figure} 

\section{EVENT GENERATION AND SIMULATION}
There are several physical processes which produce the $2$ jet + photon final
state. Our primary goal is the study of resonant $W/Z+\gamma$ production with $W/Z$ 
decaying to jets.  In addition to these signal events, we also generate 
two QCD backgrounds which contribute to the two-jet + photon events. The first 
contributor is non-resonant two-jet + photon production and the second is three-jet 
production where one of the jets is misidentified as a photon.
\par
A CDF detector simulation produces event records with the same 
structure as the data.  These generated events are then subject to the same
photon and jet selection cuts, and detector geometric acceptance, as applied to the data. 
\subsection{Signal Simulation}
\par
The signal processes $\bar{p}p \rightarrow W + \gamma$ and $\pbp \rightarrow Z 
+ \gamma$, in which the $W/Z$ decay to $2$ jets, are modeled using a leading order (LO) matrix 
element calculation \cite{baur} which includes contributions from initial and final state
inner bremsstrahlung processes. The Monte Carlo simulation
produces photons with $E_T \ge  15$~GeV  and $\Delta R$ between the
photon and partons greater than $0.4$. 
We also require that the invariant mass of the quark-antiquark pair
from the $W/Z$ decay is greater than the boson mass
minus three times the boson natural width ($\sqrt{\hat{s}}>M_V-3\Gamma_V$). This 
latter cut tends
to suppress final state bremsstrahlung and produces a
characteristic $W/Z$ resonant mass peak that simulates the
signal searched for in our data (see Section $6$).
The factorization scale
was set equal to the square of the colliding partons center
of mass energy $\sqrt{\shat}$. The MRSA$^\prime$ parton distribution
functions were used. The parton-level events are run through
HERWIG for parton shower evolution and hadronization \cite{herwig}.
The resulting  dijet mass distributions from $W$ and $Z$ decay, together with the 
combined distribution, are shown in Figure~$4$. 
\begin{figure}[htb!]
\begin{center}
\includegraphics[width=\columnwidth]{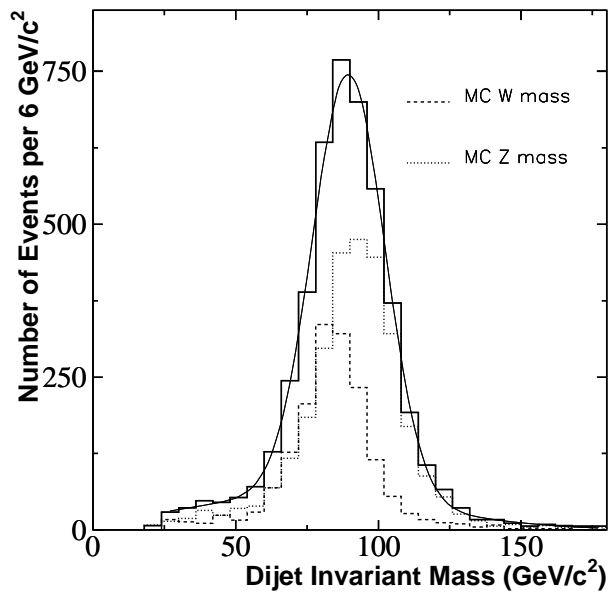}
\caption{The invariant mass distributions of the two leading jets from 
the LO $W \gamma$ and $Z \gamma$ 
simulations, together with their sum (solid line). The combined invariant mass
was fit with a double Gaussian as described in Section 5.   
\label{Figure 4}}
\end{center} 
\end{figure}
\par
The $W\gamma$ and $Z\gamma$
signal acceptance is calculated with a full CDF detector simulation. The details
of the acceptance calculations are presented in Tables $1$ and $2$ for the
$W\gamma$ and $Z\gamma$ events respectively. The uncertainty in the acceptances for each
 cut were estimated by varying the upper and lower limits on the cut
parameters and measuring the effects on the number of events
passing the cuts. The errors in the Table $1$ and Table $2$ represent cumulative errors on 
the acceptance, where all the errors
of the subsequent cuts are added in quadrature.
\begin{table}[ht!]
\caption{W$\gamma$ event selection acceptances \label{tab:w_eff}}
\begin{center}
\begin{tabular}{|l|c|c|c|}\cline{1-4}
W$\gamma$ sample& events& $\%$ & error \%\\
\cline{1-4} generated  &            97298  & 100  & \\
\cline{1-4} photon $|\eta|$ &   55487  & 57.03 & + 2.40, --2.83\\
\cline{1-4} photon ID           &   25939  & 26.67  & + 1.13, --2.02\\
\cline{1-4} photon trigger &    9567   & 9.83  & + 0.62, --0.83\\
\cline{1-4} jet $E_T$ + $|\eta|$&         1691   & 1.74  & + 0.30, --0.29 \\
\cline{1-4} extra jet&         1691   & 1.74  & + 0.31, --0.30 \\
\cline{1-4} tracks&  1677 & 1.72  & + 0.31, --0.30 \\
\cline{1-4}
\end{tabular}
\end{center}
\end{table}
\begin{table}[h!]
\caption{Z$\gamma$ event selection acceptances \label{tab:z_eff}}
\begin{center}
\begin{tabular}{|l|c|c|c|}\cline{1-4} Z$\gamma$ sample& events& $\%$ & error \%\\
\cline{1-4} generated  &                102210   & 100 & \\
\cline{1-4} photon $|\eta|$ &       68563  & 67.08  & + 2.55, --3.16\\
\cline{1-4} photon ID           &       37080  & 36.28  & + 1.38, --2.78\\
\cline{1-4} photon trigger &        15294  & 14.96  & + 0.86, --1.45\\
\cline{1-4} jet $E_T$ + $|\eta|$&   2959  & 2.90   & + 0.47, --0.49 \\
\cline{1-4} extra jet&              2959   & 2.90  & + 0.48, --0.51 \\
\cline{1-4} tracks&     2946  & 2.88   & + 0.48, --0.51 \\
\cline{1-4}
\end{tabular}
\end{center}
\end{table}
\par
The photon $\eta$ cut was varied from $1.0$ to $1.2$.  The photon ID cut combines the number of charged 
tracks associated with a photon cluster, the sum of transverse momenta $p_{T}$ of the charged tracks, and 
the photon isolation.  The number of charged tracks allowed in a photon
cluster cone was varied from zero to two. The effect
of this change was propagated through to the photon
isolation cuts, and the resulting variation in the acceptance
was used as an estimate of the systematic uncertainty in the
photon identification. Finally, we estimated the
uncertainty in the acceptance due to the photon trigger.
The key threshold points in the turn-on function~\cite{dana1}
were varied by  $\pm 5 \%$  from the nominal value. For example, 
the photons with $23$ GeV $\le E_T <$ $26$ GeV originally had $22$\% probability to pass the trigger. The 
high acceptance
cut lowers this probability to $17$\%, while the low acceptance cut raises this probability to $27$\%. 
In addition, the photons with $26$ GeV $\le E_T <$ $30$ GeV had $77$\% probability to pass the trigger. Finally,
photons with $E_T \ge 30$ GeV were passing the trigger with 97\%  probability.
The 
resulting
effect on the acceptance was taken as a systematic uncertainty.
Tables $1$ and $2$ summarize the photon selection acceptances and
their uncertainties.
\par
The systematic uncertainties due to jet selection
criteria can be divided  
into three categories: those based upon the cuts used to select the two
high $E_T$ jets
($E_T > 15$ GeV, and $|\eta|<1.1$); those caused by rejection of events  having
additional low $E_T$ jets ( additional jet
can not have $E_T>10$ GeV if within $|\eta|<2.4$); and
those due to the
use of charged tracks in the jets to define the event vertex.
The uncertainty caused by the high $E_T$ jet selection was estimated
by varying the $E_T$ cut from $14$ to $16$ GeV and the $|\eta|$ cut from
$1.0$ to $1.2$ (see "jet $E_T$ and $\eta$" rows in Tables 1 and 2). The effect of the
rejection of additional jets was evaluated by varying
$E_T$ from $9.5$ to $10.5$ GeV and $|\eta|$ from 2.3 to 2.4 ("extra jet"
rows in Tables 1 and 2). The final uncertainty on the jet selection
comes from the use of charged tracks to  determine  the event
vertex. The upper number on the tracking uncertanty was estimated by demanding
that each of the jets has at least 2 charged tracks which in addition have
very small distance from the calculated jet vertex. 
Variations in the track selection criteria 
cause almost negligable uncertainty of event acceptance shown in the last
rows of Tables 1 and 2.
The final overall event selection acceptance for  $W\gamma$ events
is $0.017\pm 0.003$ and for  $Z\gamma$ events $0.029\pm 0.005$. 
\subsection{Background Simulation}
We use a tree level calculation for the $\gamma~+~2$ jets background process which 
includes both prompt photon and 
bremsstrahlung contributions \cite{owens}, followed by HERWIG for parton evolution and 
fragmentation. In addition, three-jet events are generated using PYTHIA \cite{pythia}, with 
JETSET performing the parton evolution and fragmentation. 
Each of the three leading jets were tested for misidentification as photons. This
procedure was based on the probability distribution measured from jet data which 
describes how often a jet with a  particular $E_T$ is misidentified as a photon.
The data show that a jet with $E_T \ge 23$~GeV has a maximum 
probability of $8\times 10^{-4}$ to  fake a photon.
\par
The di-jet mass distribution from the data favors a background mixture of
$60$\% two-jet+photon and $40$\% three-jet in which one jet fakes a photon.
Figure~$6$ shows the dijet mass distribution for this combination
of backgrounds.
For masses between $60$ GeV/$c^2$ and $126$ GeV/$c^2$ the background distribution is fit well to
the form $dN/dm_{jj} =$ $ A\exp(-bm_{jj})$ as shown
by the solid line in Figure~$5$. This slope is known to be effected by the
inclusion
of higher-order QCD contributions, so we leave the exponential slope of the
background as a free parameter in the fit to the experimental dijet mass
mass spectrum to a combination of signal plus background (see Section $6$
for details).
\begin{figure}[htb!]
\begin{center}
\includegraphics[width=\columnwidth]{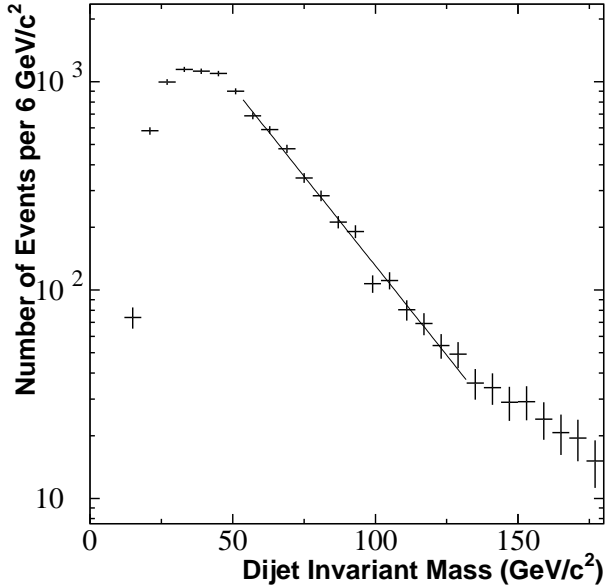}
\caption{Combined background of two-jet + $\gamma$ events and three-jet events in which one 
jet is identified as a photon. The distributions are mixed in the ratio $60\%$ to $40\%$. The exponential 
fit is presented with a solid line.
\label{Figure 5}}
\end{center}
\end{figure}
\section{STANDARD MODEL PREDICTIONS FOR $W\gamma$, $Z\gamma$ CROSS SECTIONS}
Four parton--level event samples were generated using the
procedure described in Section $4.1$: $W^{+}\gamma$ (where
$W^{+}\rightarrow u\bar{d}$),   $W^{-}\gamma$ (where
$W^{-}\rightarrow d\bar{u}$), $Z\gamma$ (where
$Z\rightarrow u\bar{u}$), and   $Z\gamma$ (where
$Z\rightarrow d\bar{d}$).  Partons are treated as massless in the model.
In order to include the $W/Z$ decays to heavy quarks, the initial ``light" cross sections
 are multiplied by appropriate factors. For example,
$W^{+}\rightarrow u\bar{d}$ is multiplied by $2.1$ to account for $c\bar{s}$, $c\bar{d}$, $u\bar{s}$
and $c\bar{b}$ decays. The same is valid for the $W^{-}\rightarrow d\bar{u}$. The cross section for
$Z\rightarrow u\bar{u}$ is multiplied by $2.0$ to account for $Z\rightarrow c\bar{c}$, while the cross section
for $Z\rightarrow d\bar{d}$ is multiplied by factor $3.0$ to include  $Z\rightarrow s\bar{s}$, $b\bar{b}$ 
decays.  In addition, the final  cross sections are multiplied by a factor of $1.3$ \cite{baurcomm}  to take 
into account higher order QCD
corrections.
\par
 The resulting cross section for $p\bar{p} \rightarrow W^{\pm}+\gamma$  with $W^{\pm}$ decay to 
quark-antiquark is $6.35$ pb, and  for $p\bar{p} \rightarrow Z+\gamma$ with the $Z$ decay to 
quark-antiquark is $6.52$ pb. These values pertain to events generated with
$E_T(\gamma) \geq$ $15$~GeV, $\Delta R$ between the photon and each of the 
quarks $\Delta R(\gamma-q)> 0.4$ and the quark-antiquark mass greater
than the boson mass minus three times the boson
natural width.
\par
With an integrated luminosity of $90$ pb$^{-1}$ we would expect 
$N_{W\gamma} = 90 \times 6.35 \times 0.017 = 10^{+2}_{-2}$ events, and 
$N_{Z\gamma} = 90 \times 6.52 \times 0.029 = 17^{+3}_{-3}$ events. The total is
 $27^{+5}_{-5}$ events in the 
two-jet + photon data sample in the dijet invariant mass region between $60$ and $126$ GeV/$c^2$. 
The errors on the total number of events were added directly, as they result from the same detector systematics
in the Monte Carlo simulation. The 
generated $W^{\pm}\gamma$ and $Z\gamma$ events are combined 
in proportion to their cross sections.
\par
The dijet mass distribution from the combined $W/Z\gamma$ generated events shown Figure~$4$ is 
normalized to a total area of one and fit to a double Gaussian 
\[ \frac{1}{N} \frac{dN}{dm_{jj}}= S(m_{jj})  \]
\[S(m_{jj}) = S_1 e^{-\frac{(m_{jj}-m_1)^2}{(2\sigma_1)^2}} +  S_2 
              e^{-\frac{(m_{jj}-m_2)^2}{(2\sigma_2)^2}}\]
\noindent The fit parameters are $S_1=0.028$ (GeV/c$^2)^{-1}$, $m_1=89.6$ GeV/$c^2$, 
$\sigma_1=12.5$ GeV/$c^2$, and $S_2=0.002$ (GeV/c$^2)^{-1}$, $m_2=76.3$ GeV/$c^2$,  
$\sigma_2=41$ GeV/$c^2$. These parameters are used in fitting the dijet mass 
distribution from the data to a combination of an exponential background and the 
signal distributions. 
\par
The widths of the invariant mass spectra for simulation of $W$ and 
$Z$ production shown in Figure~$4$ are consistent with the inferred mass resolution 
from a study of inclusive two jet production at CDF \cite{massres}. Over the dijet mass 
range from $60$ to $126$ GeV/$c^2$ the fractional dijet mass resolution ($\delta_{m_{jj}}/m_{jj}$) is 
essentially constant at $15\%$.
\section{ FITTING STANDARD MODEL PREDICTIONS TO THE DATA}
A sum of predicted background and resonant dijet mass distributions is fit to a 
histogram of the data in $11$ bins over the mass range $60$ to $126$ GeV/$c^2$.  The 
fitting procedure maximizes the likelihood function:
\[L=\exp(-\chi^2/2)\]
\noindent where the $\chi^2$ is given by
\[\chi^2(\tilde{f},\tilde{b}) = \sum_{i=1}^{n} 
\frac{(N_i-\tilde{N_i}(\tilde{f},\tilde{b}))^2}{\sigma_i^2}\] 

The optimal set of the two free parameters ($\tilde{f},\tilde{b}$) is found by 
minimizing the $\chi^2$ of the fit to the dijet mass distribution of the data.  
$N_i$ and $\sigma_i$  are the number of events in the $i^{th}$ mass bin of the 
data, and the statistical error. The total number of events in the mass range 
from $60$--$126$~GeV/$c^2$ is $N_o=2656$. The data distribution $d\tilde{N}/d m_{jj}$ is 
described by a combination of background ($B$) and resonant $W/Z$ signal ($S$).

\begin{eqnarray*}
\frac{d\tilde{N}}{d m_{jj}} & = & N_o((1-\tilde{f})B(m_{jj},\tilde{b})+\tilde{f}S(m_{jj}))
\end{eqnarray*}

Here, $\tilde{f}$ is the fraction of the signal in data and $\tilde{b}$ is the 
exponential slope of the background dijet mass distribution.  Functions 
$B(m_{jj},\tilde{b})$ and $S(m_{jj})$ represent parametrized shapes of the dijet mass spectrum for the 
background and the signal. The background distribution is represented by an exponential
\[B(m_{jj},\tilde{b}) = B_oe^{\tilde{b}m_{jj}}\]
\noindent while the signal is represented by the double Gaussian distribution function 
$S(m_{jj})$ described  above. The functions $B(m_{jj},\tilde{b})$ and $S(m_{jj})$ are normalized so that 
the area under each, over the mass range from $60$ to $126$ GeV/$c^2$, is one.
\[\int_{60}^{126} B(m_{jj},\tilde{b})dm_{jj}=1\]
\[\int_{60}^{126} S(m_{jj})dm_{jj}=1\]
The prediction for the number of events in the i$th$ mass bin is:
\[\tilde{N}_i(\tilde{f},\tilde{b})=\int_{bin~i} \frac{d\tilde{N}}{dm_{jj}} dm_{jj}\]
The cross section for production of $W\gamma$ and $Z\gamma$ ($\sigma_{W/Z\gamma}$) can be 
expressed in terms of the fraction $\tilde{f}$ 
\[\tilde{\sigma}_{W/Z\gamma}=\frac{N_o\tilde{f}}{\varepsilon_{W/Z\gamma}\mathcal{L}}\]
\noindent where $\mathcal{L}$ is the total integrated luminosity of $90 \pm 4$ 
pb$^{-1}$ \cite{lumin}, and $\varepsilon_{W/Z\gamma}$ is the combined $W/Z 
\gamma$ event detection acceptance = $0.023\pm 0.004$. This final acceptance is an average 
of   $\varepsilon_{W\gamma}$ and $\varepsilon_{Z\gamma}$ quoted in Tables $1$ and $2$. The errors on the
final acceptance are calculated as weigthed average of the errors on the individual $W\gamma$ and $Z\gamma$
acceptances.
\par
From the LO QCD predictions for the background and signal cross sections, we 
would expect about $1\%$ of the events in the fit region of the dijet invariant 
mass spectrum ($60$--$126$ GeV/$c^2$) to originate from the hadronic decay of $W/Z$ 
bosons.  The fit that maximizes the likelihood function $L(\tilde{f})$ gives 
$f$ of $-0.05 \pm 0.05$, corresponding to an unphysical negative cross section. This fit to the data is shown as 
the solid line in Figure $3$. 
\par
Since no signal is observed, we calculate upper limits on the
$W/Z\gamma$ cross section, $\sigma$, using a Bayesian procedure.
After maximizing it with respect to $\tilde{b}$, the likelihood
   function depends on the parameters $\sigma$, ${\cal L}$, and
   $\epsilon_{W/Z\gamma}$.  A posterior probability density is
   obtained by multiplying this reduced likelihood with truncated
   Gaussian prior densities for ${\cal L}$ and $\epsilon_{W/Z\gamma}$,
   with mean and width equal to the central value and uncertainty of
   these parameters, respectively.  The cross section is assigned a
   uniform prior.
The posterior density is then integrated over the
parameters $\mathcal{L}$ and $\epsilon_{W/Z\gamma}$, and the upper limit
for the $W/Z\gamma$ production cross section is obtained by calculating
the $95^{th}$ percentile of this distribution (the value of $\sigma$ such 
that 95\% of the area under this distribution is below this value). 
The posterior density integrated over $\mathcal{L}$ and 
$\epsilon_{W/Z\gamma}$ as a function of $\sigma$ is shown in Figure $6$. 
\begin{figure}[htb!]
\includegraphics[width=\columnwidth]{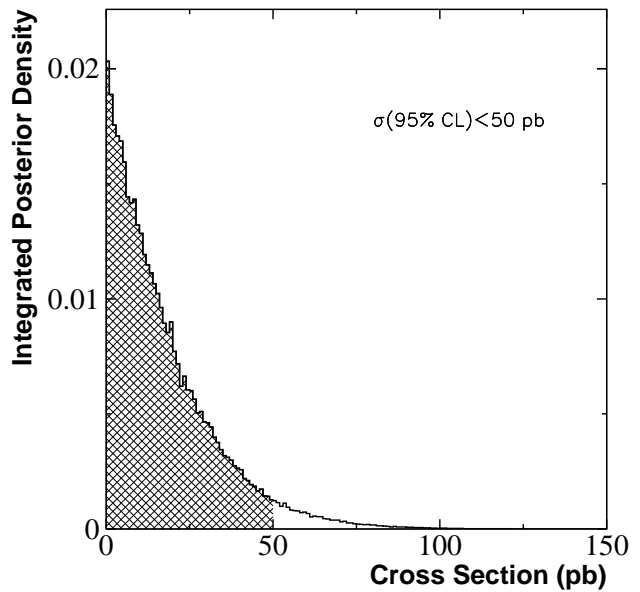}
\caption{Posterior density (Likelihood function multiplied by Gaussian 
   priors for $\mathcal{L}$ and $\epsilon_{W/Z\gamma}$) integrated over 
   $\mathcal{L}$ and $\epsilon_{W/Z\gamma}$, versus $\sigma$. 
\label{Figure 6}}
\end{figure}
The result is  
$\sigma(\bar{p}p\rightarrow W\gamma)\times BR(W\rightarrow jj)$ + 
$\sigma(\bar{p}p\rightarrow Z\gamma)\times BR(Z\rightarrow jj) \le 50~$pb 
 compared to a Standard Model expectation of 13 pb.
\section{SUMMARY}
We have searched for the process $\pbp \rightarrow W/Z + \gamma + X$ in which 
the $W$ and $Z$ bosons decay hadronically.  The sensitivity of the search is 
enhanced by fitting the dijet mass spectrum from $\pbp \rightarrow 2$ jet $+ 
\gamma$ to that expected from QCD models for the background and the line shape from the $W/Z \rightarrow 2$ jets decay.
The Standard Model prediction of the cross section for
$\pbp \rightarrow W/Z + \gamma$, with the $W/Z$ bosons decaying to quark-
antiquark pairs, is $13$ pb. This cross section is for photons with
of $E_T(\gamma)>15$ GeV,  $\Delta R(\gamma-q) >0.4$ and 
for $\sqrt{\hat{s}}>M_V-3\Gamma_V$.
\par
The data from $90$ pb$^{-1}$ of $\pbp$ interactions show no evidence of 
this process. By integrating the likelihood function, we get a $95\%$ confidence level upper limit 
to the cross section of $50$ pb.  This analysis method could prove effective for measuring 
the $W/Z \rightarrow$ dijet mass signal in this channel from a larger statistics sample of data 
being collected by the CDF experiment in Tevatron Run $2$.
\par
We thank the Fermilab staff and the technical staffs of the participating
institutions for their vital contributions. We also thank Uli Baur and Jeff 
Owens for valuable discussions.
This work was supported by the U.S. Department of Energy and National Science Foundation; the Italian 
Istituto Nazionale di Fisica Nucleare; the Ministry of Education, Culture, Sports, Science and Technology 
of Japan; the Natural Sciences and Engineering Research Council of Canada; the National Science Council 
of the Republic of China; the Swiss National Science Foundation; the A.P. Sloan Foundation; the 
Bundesministerium fuer Bildung und Forschung, Germany;
the Korean Science and Engineering Foundation and the Korean Research 
Foundation; the Particle Physics and Astronomy Research Council and the Royal Society, UK; the 
Russian Foundation for Basic Research; the
Comision Interministerial de Ciencia y Tecnologia, Spain; and in part by the
European Community's Human Potential Programme under contract HPRN-CT-20002, 
Probe for New Physics.

\end{document}